\documentclass[aps,twocolumn,groupedaddress,showpacs,floats]{revtex4}
\usepackage[dvips]{graphicx}
\usepackage{bm}
\begin{document}

\title{Kelvin Wave Cascade and Decay of Superfluid Turbulence}

\author{Evgeny Kozik}
\author{Boris Svistunov}
%\thanks{}

\affiliation{Department of Physics, University of
             Massachusetts, Amherst, MA 01003}
\affiliation{Russian Research Center ``Kurchatov Institute'',
123182 Moscow }

\begin{abstract}
Kelvin waves (kelvons)---the distortion waves on vortex
lines---play a key part in the relaxation of superfluid turbulence
at low temperatures. We present a weak-turbulence theory of
kelvons. We show that non-trivial kinetics arises only beyond the
local-induction approximation and is governed by three-kelvon
collisions; corresponding kinetic equation is derived. On the
basis of the kinetic equation, we prove the existence of
Kolmogorov cascade and find its spectrum. The qualitative analysis
is corroborated by numeric study of the kinetic equation. The
application of the results to the theory of superfluid turbulence
is discussed.
\end{abstract}


% 67.40.Vs Vortices and turbulence

% 03.75.Lm Tunneling, Josephson effect, Bose-Einstein condensates in
%          periodic potentials, solitons, vortices and topological
%          excitations

% 47.32.Cc Vortex dynamics

\maketitle

The distortion waves on a vortex filament---Kelvin waves
(KW)---have been known for more than a century \cite{Kelvin}.
Superfluids with their topological (quantized) vorticity form a
natural domain for KW \cite{Donnelly}.  Nowadays there is a strong
interest to the non-linear aspects of KW associated with studying
low-temperature superfluid turbulence of $^4$He
\cite{Sv95,Vinen2000,Davis,Tsubota,Kivotides,Vinen2001,Vinen2002,Vinen2003},
as well as vortex dynamics in ultra-cold atomic gases
\cite{Bretin,Mizushima}.

The superfluid turbulence \cite{Donnelly,Cambridge} is a chaotic
tangle of vortex lines. In the absence of the normal component ($T
\to 0$ limit), KW play a crucial part in the vortex tangle
relaxational dynamics. In contrast to a normal fluid, the
quantization of the velocity circulation in a superfluid makes it
impossible for the vortex line to relax by gradually slowing down.
The only allowed way of relaxation is reducing the total line
length. At $T=0$ even this generic scenario becomes non-trivial,
as the total line length is, to a very good approximation, a
constant of motion. In the scenario proposed by one of us
\cite{Sv95}, the vortex line length---in the form of KW generated
in the process of vortex line reconnections---{\it cascades} from
the main length scale (typical interline separation, $R_0$) to
essentially lower length scales; ultimately decaying into phonons,
as it was pointed out by Vinen \cite{Vinen2000,Vinen2001}.

A very specific feature of KW cascade is that the intrinsic vortex
line dynamics in the local-induction approximation (LIA) (for an
introduction, see, e.g., \cite{Donnelly,Cambridge}) controlled by
the small parameter $1/\ln(R_0/\xi)$, with $\xi$ the vortex core
radius, is subject to a specific curvature-conservation constraint
rendering it unable to support the cascade process \cite{Sv95}
(see also below). Within LIA, an ``external" ingredient of the
vortex line dynamics---the vortex line crossings with subsequent
reconnections---is required to push the KW cascade down towards
arbitrarily small wavelengths. The most characteristic feature of
this LIA scenario, distinguishing it from typical non-linear
cascades, is the fragmentation of the vortex lines due to local
self-crossings \cite{Sv95}; we will thus refer to this scenario as
{\it fragmentational} scenario.

Experimentally, the main consequence of the existence of a cascade
regime, no matter what is its microscopic nature, is independence
of the relaxation time of superfluid turbulence on temperature in
the $T \to 0$ limit. Davis {\it et al}. have observed such a
regime set in in $^4$He at $T<70mK$ \cite{Davis}. The numeric
simulation of the vortex tangle decay at $T=0$ performed by
Tsubota {\it et al}. \cite{Tsubota} within the framework of LIA
has clearly revealed the cascade regime; this may be considered as
a circumstantial \cite{rem} evidence of the proposed in
\cite{Sv95} scenario.

A general question arises, however, of how far, in the wavenumber
space, the structure of KW cascade is pre-determined by LIA
dynamics. At wavelengths  $\lambda \ll R_0$ the non-local effects
of the vortex line dynamics compete with LIA dynamics and can
ultimately become the main driving force of the cascade. Moreover,
given the specific spectrum of KW turbulence associated with the
fragmentational scenario, where the amplitude of the turbulence is
smaller than the wavelength only by a logarithmic factor
\cite{Sv95}, one concludes, that if non-local effects can in
principle support the cascade, no matter how small is the
corresponding contribution at the main wavelength scale $R_0$,
there will inevitably be such a wavelength scale $\lambda_* \ll
R_0$, where the fragmentational scenario will be replaced by a
purely non-linear---to be referred as {\it pure}---scenario in
which vortex line self-crossings play no role. The existence of
the crossover between the two cascade scenarios has at least two
important implications. First, the spectrum of sizes of vortex
rings generated by decaying superfluid turbulence will have a
lower cutoff $\sim \lambda_*$. Secondly, the spectrum of KW
turbulence will be changed, which, in particular, is crucial for
the cascade cutoff theory \cite{Vinen2001} where the
characteristic wavelength $\lambda_{\rm ph}$ at which KW
essentially decay into phonons is a function of the cascade
spectrum.

A strong numeric evidence in favor of existence of pure KW cascade
has been reported by Kivotides {\it et al}. \cite{Kivotides}.
Being very expensive numerically, this simulation was not able to
accurately resolve the spectrum of KW turbulence. The data of
recent simulation by Vinen {\it et al}. \cite{Vinen2003} seem to
be more conclusive.

In this Letter, we propose a self-consistent (asymptotically exact
in the limit of high wavenumbers) treatment for the pure KW
cascade. More generally, we develop the KW kinetic theory in the
regime of weak turbulence, where the smallness of non-linearities
reduces the non-linear effects to scattering processes for the
harmonic modes. We find that the leading elementary process
responsible for the kinetics is the three-kelvon scattering. We
demonstrate that the kinetics is {\it entirely} due to the
spatially non-local interactions, the local contributions exactly
cancelling each other. This cancellation sets the limit on the
maximum power of the pure cascade: At the wavelength scale $\sim
R_0$ the contribution of the kelvon-scattering processes to the KW
cascade contains a small factor $1/\ln^2(R_0/\xi)$ as compared to
the reconnection-induced part.

In terms of weak turbulence theory, KW cascade is a Kolmogorov
cascade \cite{Zakharov}, associated with the transport of energy
(closely related to the vortex line length in our case) in the
wavenumber space. We establish the Kolmogorov spectrum and find
the relation between the energy flux and the amplitude of KW
turbulence. Finally, on the basis of our results we discuss the
fragmentational-to-pure cascade crossover, which we predict to be
rather extended (up to two decades) in the wavenumber space.

{\it Hamiltonian}. We employ the Hamiltonian representation of the
vortex line motion \cite{Sv95}, which is exact up to a certain
geometric constraint: there should exist some axis $z$, with
respect to which the position of the line can be specified in the
parametric form $x=x(z)$, $y=y(z)$, where $x$ and $y$ are
single-valued functions of the coordinate $z$. This perfectly
suits our purposes, since we are interested in the wavelengths
substantially smaller than $R_0$ and thus will treat the vortex
line as a straight line with small-amplitude distortions. In terms
of the complex canonic variable $w(z,t)=x(z,t)+iy(z,t)$, the
Biot-Savart dynamic equation acquires the Hamiltonian form
$i\dot{w}=\delta H[w] / \delta w^*$, with \cite{Sv95}
\begin{equation}
H=\frac{\kappa}{4\pi} \int \! \! dz_1 dz_2 { 1+{\rm Re} \,
w'^*(z_1)w'(z_2) \over \sqrt{(z_1-z_2)^2+|w(z_1)-w(z_2)|^2}} \; ,
\label{HAM}
\end{equation}
where $\kappa=2\pi \hbar /m$ is the circulation quantum ($m$ is
the particle mass). Our fundamental requirement that the amplitude
of KW turbulence be small as compared to the wavelength is
formulated as
\begin{equation}
\alpha(z_1,z_2) = {|w(z_1)-w(z_2)| \over |z_1-z_2|} \ll 1 \; .
\label{ALPHA}
\end{equation}
This allows us to expand (\ref{HAM}) in powers of $\alpha$:
$H=E_0+H_0+H_1+H_2+ \ldots $ ($E_0$ is just a number and will be
ignored). The terms that prove relevant are:
\begin{equation}
H_0=\frac{\kappa }{8\pi} \int {dz_1 dz_2 \over |z_1-z_2|} [2{\rm
Re} \, w'^*(z_1)w'(z_2)- \alpha^2  ] \; , \label{H0}
\end{equation}

\begin{equation}
H_1=\frac{\kappa }{32\pi} \int {dz_1 dz_2 \over |z_1-z_2|} [ 3
\alpha^4 - 4 \alpha^2 {\rm Re} \, w'^*(z_1)w'(z_2) ] \; ,
\label{H1}
\end{equation}

\begin{equation}
H_2=\frac{\kappa }{64\pi} \int {dz_1 dz_2 \over |z_1-z_2|} [ 6
\alpha^4 {\rm Re} \, w'^*(z_1)w'(z_2) - 5 \alpha^6 ] \; .
\label{H2}
\end{equation}

The Hamiltonian $H_0$ describes the linear properties of KW. It is
diagonalized by the Fourier transformation $w(z) = L^{-1/2} \sum_k
a_k e^{ikz}$ ($L$ is the system size, periodic boundary conditions
are assumed):
\begin{equation}
 H_0=\frac{\kappa }{4\pi} \sum_k    \omega_k \, a_k^* a_k \; , ~~~~
 \omega_k = (\kappa / 4\pi) \,\ln ( 1/k \xi) \, k^2  \; ,
\end{equation}
yielding Kelvin's dispersion law $\omega_k$ .

Though the problem of KW cascade generated by decaying superfluid
turbulence is purely classical, it is convenient to approach it
quantum mechanically---by introducing KW quanta, kelvons. In
accordance with the canonical quantization procedure, we
understand $a_k$ as the annihilation operator of the kelvon with
momentum $k$ and correspondingly treat $w(z)$ as a quantum field.
A minor caveat is in order here. The Hamiltonian functional
(\ref{HAM}) is {\it proportional} to the energy---with the
coefficient $\kappa \rho/2$, where $\rho$ is the mass
density---but not {\it equal} to it. This means that if one
prefers to work with genuine Quantum Mechanics rather than a fake
one (for our purposes, the latter is also enough), using true
kelvon annihilation operators, $\hat{a}_k$, field operator
$\hat{w}$, and Hamiltonian, $\hat{H}$, he should take into account
proper dimensional coefficients: $\hat{H}=(\kappa \rho/2)H$,
$\hat{w}(z)=\sqrt{{2 \hbar / \kappa \rho L}} \sum_k \hat{a}_k
e^{ikz}$. By choosing the units $\hbar = \kappa = 1$, $\rho = 2$,
we ignore these coefficients until the final answers are obtained.

In the quantum approach, there naturally arises the notion of the
number of kelvons. This number is {\it conserved} by the
Hamiltonian (\ref{HAM}) in view of its global $U(1)$ symmetry, $w
\to e^{i\varphi} w$ (reflecting rotational symmetry of the
problem). Another advantage of the quantum language in
weak-turbulence problems is that the collision term of the kinetic
equation immediately follows from the Golden Rule for
corresponding elementary processes.

{\it Kelvon scattering.} As was mentioned above, Hamiltonian
(\ref{HAM}) implies only elastic scattering. The one-dimensional
character of the problem in combination with the conservation of
the momentum and energy suppresses the two-kelvon scattering
channel: the process $(k_1,k_2) \to (k_3,k_4)$ is only possible if
either $(k_3=k_1,~k_4=k_2)$, or $(k_3=k_2,~k_4=k_1)$ which does
not lead to any kinetics. We thus conclude that the leading
process in our case is the three-kelvon scattering. The processes
involving four and more kelvons a much weaker due to the
non-equality (\ref{ALPHA}). The effective vertex,
$V_{1,2,3}^{4,5,6}$, for the three-kelvon scattering process
[subscripts (superscripts) stand for the initial (final) momenta
of the three kelvons; we use a short-hand notation replacing each
momentum $k_j$ with its index $j$] consists of two different
parts. The first part involves terms generated by the two-kelvon
vertex, $A$ (corresponding to the Hamiltonian $H_1$) in the second
order of perturbation theory. [All these terms are similar to each
other; we explicitly specify just one of them: $A_{1,2}^{4,7} \,
G(\omega_7,k_7) \, A_{7,3}^{5,6}$. Here $G(\omega,k)=1/(\omega
-\omega_k)$ is the free-kelvon propagator,
$\omega_7=\omega_1+\omega_2-\omega_4$, $k_7=k_1+k_2-k_4$.] The
second part of the vertex $V$ is the bare three-kelvon vertex,
$B$, associated with the Hamiltonian $H_2$.

The explicit expressions for the bare vertices directly follow
from (\ref{H1}) and (\ref{H2}): $A=(6D -E)/8\pi$,
$D^{4,5,6}_{1,2,3}=\int_{0}^{L} ({\rm d}x /x^5) [ \, 1 - C_{\{ 1\}
} - C_{\{ 2\} }- C^{\{ 3\} } -C^{\{ 4\} } + C^{\{ 3\} }_{\{ 2\} }
+ C^{\{ 43\} } + C^{\{ 4\} }_{\{ 2\} } ] $,
$E^{4,5,6}_{1,2,3}=\int_{0}^{L}  ({\rm d}x /x^3) [ \,k_4k_1(\:
C^{\{ 4\} }+C_{\{ 1\} } -C^{\{ 43\} }-C^{\{ 4\} }_{\{ 2\} }\:)
+k_3k_1(\: C^{\{ 3 \} } + C_{\{ 1\} } - C^{\{ 43\} } - C^{\{ 3\}
}_{\{ 2\} }
 \:) +k_3k_2(\: C^{\{ 3\} }+ C_{\{ 2\} }  - C^{\{ 43\} } -   C^{\{ 3\} }_{\{
1\} }  \:) + k_4k_2(\: C^{\{ 4\} } +C_{\{ 2\} }- C^{\{ 43\}
}-C^{\{ 3\} }_{\{ 2\} }\, ] $; $B=(3P-5Q)/4\pi$,
$P^{4,5,6}_{1,2,3}=\int_{0}^{L} ({\rm d}x /x^5) \;k_6k_2[C_{\{ 2
\} } -C^{\{ 5 \} }_{\{ 2  \} } -C_{\{  23 \} } +C^{\{ 5 \} }_{\{
23 \} } -C^{\{ 4 \} }_{\{ 2  \} } +C^{\{  45 \} }_{\{ 2 \} }
+C^{\{ 4 \} }_{\{ 23 \} } -C^{\{ 6 \} }_{\{1 \} } +C^{\{ 6 \} }
-C^{\{56 \} } -C^{\{ 6 \} }_{\{3   \} } +C^{\{ 56 \} }_{\{3 \} }
-C^{\{  46 \} } +C^{\{456 \} } +C^{\{46 \} }_{\{ 3 \} } -C_{\{ 12
\} } ] $, $Q^{4,5,6}_{1,2,3}=\int_{0}^{L} ({\rm d}x /x^7) \: [1
-C^{\{ 4  \} } -C_{\{ 1 \} } +C^{\{ 4  \} }_{\{ 1  \} } -C^{\{  6
\} } +C^{\{ 46 \} } +C^{\{ 6 \} }_{\{ 1  \} } -C^{\{ 46  \} }_{\{
1  \} } -C^{\{ 5 \} } +C^{\{ 45  \} } +C^{\{ 5  \} }_{\{ 1  \} }
-C^{\{  45 \} }_{\{ 1 \} } +C^{\{ 65  \} } -C^{\{ 456  \} } -C^{\{
56  \} }_{\{ 1  \} } +C_{\{ 23  \} } -C_{\{ 3  \} } +C^{\{ 4  \}
}_{\{ 3  \} } +C_{\{ 13 \} } -C^{\{ 4  \} }_{\{ 13 \} } +C^{\{ 6
\} }_{\{ 3 \} } -C^{\{  46 \} }_{\{  3 \} } -C^{\{ 6  \} }_{\{ 13
\} } +C^{\{ 5 \} }_{\{  2 \} } +C^{\{ 5  \} }_{\{ 3 \} } -C^{\{
45 \} }_{\{ 3 \} } -C^{\{  5 \} }_{\{ 13  \} } +C^{\{  6 \} }_{\{
2 \} } -C^{\{ 65 \} }_{\{ 3  \} } +C_{\{ 12  \} } +C^{\{ 4  \}
}_{\{ 2  \} } -C_{\{ 2  \} } ] $. Here $C$'s denote similar
looking cosine functions: $C_{\{ 1\} } = \cos k_1 x$, $C^{\{ 4 \}
}_{\{ 1  \} } = \cos (k_4 - k_1) x$, $C^{\{ 4 5\} }_{\{ 1  \} } =
\cos (k_4 +k_5 - k_1) x$, $C^{\{ 45 \} }_{\{ 12 \} } = \cos (k_4
+k_5 - k_1-k_2) x$, and so forth. The integrals for bare vertices
$A$ and $B$ logarithmically diverge at their lower limits and,
generally speaking, one has to introduce a cut-off parameter $\sim
\xi$. However, in the final answer for the effective vertex $V$
the logarithmic in $\xi$ terms cancel each other (it is important
here to take into account that, in the propagator $G$, the
frequency $\omega_k$ also diverges logarithmically with $\xi \to
0$), so that $V$ is finite in the limit $\xi \to 0$. This fact is
not just a mere coincidence. The small-$x$ contributions to the
integrals correspond to the local-induction dynamics, which cannot
lead to any kinetics because of the extra constants of motion
\cite{Sv95}. Cancellation of the LIA contributions at the level of
the effective scattering amplitude is an explicit demonstration of
this circumstance.

{\it Kinetic equation.} The kinetic equation is written in terms
of averaged over the statistical ensemble kelvon occupation
numbers $n_k = \langle a^{\dagger}_k a_k \rangle $:
\begin{equation}
\dot{n}_1 = \frac{1}{(3-1)! \; 3!} \sum_{k_2, \ldots , k_6} \left(
W_{4,5,6}^{1,2,3}-W_{1,2,3}^{4,5,6} \right) \; . \label{KE_gen}
\end{equation}
Here $W_{1,2,3}^{4,5,6}$ is the probability per unit time for the
elementary scattering event $(k_1,k_2,k_3) \to (k_4,k_5,k_6)$;
combinatorial factor compensates multiple counting the same
scattering event. For our interaction Hamiltonian $H_{\rm int} =
\sum_{k_1, \ldots , k_6} \delta(\Delta k) \,
\tilde{V}_{1,2,3}^{4,5,6} \, a^{\dagger}_6 a^{\dagger}_5
a^{\dagger}_4 a^{ }_3 a^{ }_2 a^{ }_1 $ [where the vertex
$\tilde{V}$ is obtained from $V$ by symmetrization with respect to
corresponding momenta permutations; $\delta(k)$ is understood
discretely as $\delta_{k, 0}$, and  $\Delta k =
k_1+k_2+k_3-k_4-k_5-k_6$] the Golden Rule reads:
$W_{1,2,3}^{4,5,6}=2 \pi  |(3!)^2 \tilde{V}_{1,2,3}^{4,5,6}|^2
f_{1,2,3}^{4,5,6} \, \delta(\Delta \omega) \delta(\Delta k)$,
where $f_{1,2,3}^{4,5,6}=n_1n_2n_3(n_4+1)(n_5+1)(n_6+1)$, $\Delta
\omega = \omega_1+\omega_2+\omega_3-\omega_4-\omega_5-\omega_6$.
Combinatorial factor $(3!)^2$ takes into account addition of
equivalent amplitudes. The classical-field limit of the quantum
kinetic equation (\ref{KE_gen}) is obtained by retaining only the
largest in occupation numbers terms:
\begin{equation}
\dot{n}_1 \! \! =  \!   216 \pi \! \! \!  \! \sum_{k_2, \ldots ,
k_6} \! \! \!  \! |\tilde{V}_{1,2,3}^{4,5,6}|^2  \, \delta(\Delta
\omega) \, \delta(\Delta k) \left( \tilde{f}_{4,5,6}^{1,2,3} \! -
\! \tilde{f}_{1,2,3}^{4,5,6} \right )  , \label{KE}
\end{equation}
where $\tilde{f}_{1,2,3}^{4,5,6}=n_1n_2n_3(n_4n_5+n_4n_6+n_5n_6)$.

{\it Kolmogorov cascade}. Kinetic equation (\ref{KE}) supports
Kolmogorov energy cascade \cite{Zakharov}, provided two conditions
are met: (i) the kinetic time is getting progressively smaller
(vanishes) in the limit of large wavenumbers, (ii) the collision
term is local in the wavenumber space---not to be confused with
the local-induction approximation in the real space,---that is the
relevant scattering events are only those where all the kelvon
momenta are of the same order of magnitude. We make sure that both
conditions are satisfied in our case: the condition (i) can be
checked by a dimensional estimate, provided (ii) is true. The
condition (ii) is verified numerically.

The cascade spectrum can be established by dimensional analysis of
the kinetic equation. The estimate of Eq.~(\ref{KE}) yields:
$\dot{n}_k \sim k^5 \cdot |V|^2 \cdot \omega_k^{-1} \cdot k^{-1}
\cdot n_k^5$, the factors go in the order of the appearance of
corresponding terms in (\ref{KE}). At $k_1 \sim \ldots \sim k_6
\sim k$ we have $|V| \sim k^6$ and
\begin{equation}
\dot{n}_k \;  \sim \; \omega^{-1}_k \, n^5_k \, k^{16} \;
.\label{n_dot}
\end{equation}
The energy flux (per unit vortex line length), $\theta_k$, at the
momentum scale $k$ is defined as
\begin{equation}
\theta_k \; = \;  L^{-1} \sum_{k' < k } \, \omega_{k'} \,
\dot{n}_{k'} \; , \label{flux_def}
\end{equation}
implying the estimate $\theta_k \sim k \dot{n}_k \omega_k $.
Combined with (\ref{n_dot}), this yields $\theta_k \sim n_k^5
k^{17}$, and the cascade requirement that $\theta_k $ be actually
$k$-independent leads to the spectrum:
\begin{equation}
n_k = A \, k^{-17/5} \; . \label{spect}
\end{equation}
The value of the parameter $A$ in (\ref{spect}) corresponds to the
value of the energy flux. An accurate relation between $\theta$
and $A$ is (we restore all dimensional parameters):
\begin{equation}
\theta \approx 3 \cdot 10^{-4} \,  \frac{\hbar^5 A^5}{\kappa^2
\rho^4} \; . \label{rel}
\end{equation}
The dimensionless coefficient in this formula has been established
numerically (see below) with the error $\sim 75 \%$. One should
not be confused with the appearance of the Plank's constant in
this relation---it is entirely due to the fact that we use quantum
mechanical definition of $n_k$.

The exponent $17/5$ is in a perfect agreement (within the error
bars) with the spectrum observed in Ref.~\cite{Vinen2003}
($\bar{\zeta_k^2}$ of Ref.~\cite{Vinen2003} is proportional to our
$n_k$); it fits the data much better than the exponent $3$ used by
the authors.

It is useful to express the KW cascade spectrum in terms of a
geometric characteristic---typical amplitude, $b_k$, of the KW
turbulence at the wavevector $\sim k$. By the definition of the
field $\hat{w}(z)$ we have: $b^2_k \sim L^{-1} \sum_{q \sim k}
\langle \hat{a}^{\dagger}_q \hat{a}_q \rangle = L^{-1} \sum_{q
\sim k} n_q \sim k \, n_k$. Hence,
\begin{equation}
b_k \propto A^{1/2} k^{-6/5} \; . \label{spect2}
\end{equation}
One can convert (\ref{spect}) into the curvature spectrum. For the
curvature ${\bf c}(\zeta)=
\partial^{2} {\bf s} /
\partial \zeta^{2}$ [where ${\bf s}(\zeta)$
is the radius-vector of the curve as a function of the arc length
$\zeta$], the spectrum is defined as Fourier decomposition of the
integral $I_c=\int |{\bf c}(\zeta)|^2 \, {\rm d} \zeta $. The
smallness of $\alpha$, Eq.~(\ref{ALPHA}), allows one to write $I_c
\approx \int {\rm d}z \langle \hat{w}''^{\dagger}(z) \hat{w}''(z)
\rangle = \sum_k k^4 \, n_k \propto \sum_k  k^{3/5}$, arriving
thus at the exponent $3/5$.

{\it Numerics.} The aim of our numeric analysis is (i) to make
sure that the collision term of the kinetic equation is local and
(ii) to establish the value of the dimensionless coefficient in
(\ref{rel}). The analysis is based on the following reasoning
\cite{Sv91}. Consider a power-law distribution of occupation
numbers, $n_k=A / k^{\beta}$, with the exponent $\beta$
arbitrarily close, but not equal, to the cascade exponent $\beta_0
=17/5$. Substitute this distribution in the collision term of the
kinetic equation---right-hand side of (\ref{KE}). Given the scale
invariance of the power-law distribution, the following
alternative takes place. Case (1): collision integral converges
for $\beta$'s close to $\beta_0$, and, in accordance with a
straightforward dimensional analysis, is equal to
\begin{equation}
{\rm Coll}([n_k=A/k^{\beta}],k) = C(\beta) A^5 \omega^{-1}_k /
k^{5\beta - 16} \; . \label{coll}
\end{equation}
Here $C(\beta)$ is a dimensionless function of $\beta$, such that
$C(\beta_0)=0$ since the cascade is a steady-state solution. Case
(2): collision integral diverges for $\beta$ close to $\beta_0$.
The case (2) means that the collision term is non-local and the
whole analysis in terms of Kolmogorov cascade is irrelevant.
[Fortunately, our numerics show that we are dealing with the case
(1).] Substituting (\ref{coll}) for $\dot{n}_k$ in
(\ref{flux_def}), we obtain the expression $\theta_k(\beta)\, = \,
(A^5/2\pi) \,  k ^{17- 5\beta} \, C(\beta)/(17- 5\beta)$. Taking
the limit $\beta \to \beta_0$, we arrive at the $k$-independent
flux $\theta = - C'(\beta_0) A^5 /10 \pi $. We use this formula to
obtain the coefficient in (\ref{rel}) by calculating $C(\beta)$
and finding its derivative $C'(\beta_0)$. We simulated the
collision integral by Monte Carlo method. The integrals $D,E,P,Q$,
were calculated numerically. The slowing down of the simulation
due to the latter circumstance is the main source of large
relative error.

{\it Superfluid turbulence.} The main characteristic of the
superfluid turbulence is the vortex line density, $L$ (total line
length per unit volume; note that $R_0 \sim 1/\sqrt{L}$.) The
parameter $A$ is related to $L$ through the energy flux $\theta$.
Since (to a good accuracy) the energy is the total line length
times $(\rho \kappa^2 /4\pi) \ln (R_0/ \xi)$, the same coefficient
relates the energy flux to the line-length flux. The line-length
flux, $\dot{L}/L$, is available from the simulation of
Ref.~\cite{Tsubota}: $\dot{L}/L \approx \, 6 \cdot 10^{-3} \kappa
\, L \, \ln(1/\xi \sqrt{L})$. We thus get
\begin{equation}
(\hbar A/\kappa \rho)^5 \approx \, 2 \, L \, \ln^2(1/\xi \sqrt{L})
\; .
\end{equation}
For the amplitude spectrum this yields:
\begin{equation}
b_kk \sim \left[ (\sqrt{L}/k) \, \ln (1/\xi \sqrt{L})
\right]^{1/5}
 \; . \label{est}
\end{equation}
With this spectrum, Vinen's prediction for the cutoff momentum
$k_{\rm ph} \sim \lambda_{\rm ph}^{-1}$, based on Eq.~(2.24) of
Ref.~\cite{Vinen2001}, should read ($c$ is the sound velocity):
\begin{equation}
k_{\rm ph}/\sqrt{L} \sim {\ln^{4/9} (1/\xi \sqrt{L}) \over
\ln^{10/9} (1/\xi k_{\rm ph}) }\left[{c \over \kappa \sqrt{L} }
\right]^{5/6}
 \; . \label{sound}
\end{equation}

From (\ref{est}) it is seen that the crossover from the
fragmentational cascade regime to the pure one should be very
slow. Indeed, to significantly suppress the fragmentational
regime---that is to suppress local self-crossings of the vortex
line---it is necessary to make the parameter $b_k k$ substantially
smaller than unity. In view of the exponent $1/5$, this requires
increasing $k$ by $\sim$ two orders of magnitude. In this extended
crossover region, one has $b_k k \sim 1$, while for the
applicability of the treatment developed in this Letter one needs
$b_k k \ll 1$. Hence, a reliable description of the crossover from
the fragmentational to the pure cascade seems to be impossible
without a direct numeric simulation of vortex line dynamics.

We are grateful to S.~Nazarenko, W.~Vinen, C.~Barenghi, and
M.~Tsubota for useful discussions.

\end{document}